# High-pressure synthesis and spin glass behavior of a Mn/Ir disordered quadruple perovskite $CaCu_3Mn_2Ir_2O_{12}$

Xubin Ye[1,2], Zhehong Liu[1,2], Weipeng Wang[1,2], Zhiwei Hu[3], Hong-Ji Lin[4], Shih-Chang Weng[4], Chien-Te Chen[4], Richeng Yu[1,2], Liu-Hao Tjeng[3], Youwen Long[1,2,5]

[1]Beijing National Laboratory for Condensed Matter Physics, Institute of Physics, Chinese Academy of Sciences, Beijing 100190, China
[2]School of Physical Sciences, University of Chinese Academy of Sciences, Beijing 100049, China
[3]Max Planck Institute for Chemical Physics of Solids, Dresden 01187, Germany
[4]National Synchrotron Radiation Research Center, Hsinchu 30076, Taiwan, ROC
[5]Songshan Lake Materials Laboratory, Dongguan, Guangdong 523808, China

E-mail: ywlong@iphy.ac.cn

**Abstract**

A new $3d$-$5d$ hybridized quadruple perovskite oxide, $CaCu_3Mn_2Ir_2O_{12}$, was synthesized by high-pressure and high-temperature methods. The Rietveld structure analysis reveals that the compound crystallizes in an $AA'_3B_4O_{12}$-type perovskite structure with space group $Im$-3, where the Ca and Cu are 1:3 ordered at fixed atomic positions. At the B site the $3d$ Mn and the $5d$ Ir ions are disorderly distributed due to a rare equal +4 charge states for both of them as determined by X-ray absorption spectroscopy. The competing antiferromagnetic and ferromagnetic interactions among $Cu^{2+}$, $Mn^{4+}$, and $Ir^{4+}$ ions give rise to spin glass behavior, which follows a conventional dynamical slowing down model.

## 1. Introduction

The compounds with $ABO_3$ perovskite structure composed of a 3D framework of corner-sharing $BO_6$ octahedra have attracted much attention, due to their large varieties of physical and chemical properties [1-5]. Chemical substitution either at $A$ or $B$ site is an effective method to tune the structure and physical properties of perovskites. For example, the $A$-site alkaline earth substitution for rare earth in manganates $La_{1-x}A_xMnO_3$ leads to insulator-to-metal and paramagnetism-to-ferromagnetism transitions [6-8]. The $B$-site substitution of $Fe^{4+}$ by $Co^{4+}$ in $SrFe_{1-x}Co_xO_3$ causes a series of magnetic transitions [9]. Specifically, as the $x$ increases, the magnetic ground state changes from helimagnetic antiferromagnetic (AFM) state through cluster glass to ferromagnetic (FM) state with a high Curie temperature. In particular, if three quarters of $A$ site in a simple $ABO_3$ is substituted by a transition metal ion $A'$, one may obtain an A-site ordered quadruple perovskite with the chemical formula of $AA'_3B_4O_{12}$ [10-14]. Since the ionic size of a transition metal is much less than that of a typical $A$-site cation such as alkaline, alkaline earth or rare earth metal, to sustain the perovskite structure, square-coordinated $A'O_4$ units and heavily tilted $BO_6$ octahedra show up in $AA'_3B_4O_{12}$, as presented in figure 1(a). The peculiar coordination environment of $A'$ site is favorable to accommodate a Jahn-Teller distortion ion like $Mn^{3+}$ or $Cu^{2+}$. As a $Cu^{2+}$ ion with $3d^9$ configuration is concerned, this square coordination may result in a $d_{xy}$ hole state with the Cu-O square planar oxygen bond pointing between $x$ and $y$ [15]. The electronic states of $Cu^{2+}$ can exhibit either localized or itinerant behavior depending on the electronic properties of the $B$-site ions. Moreover, the $A'$-site $Cu^{2+}$ spins can form a magnetic lattice with face-centered cubic symmetry. The direct-exchange interactions between the nearest neighboring $Cu^{2+}$ ions compete with the superexchange interactions caused by Cu-O-$B$ pathways, leading to unusual magnetic instability and even a magnetic tri-critical point [16-18].

CaCu$_3$Mn$_4$O$_{12}$ and CaCu$_3$Ir$_4$O$_{12}$ are two interesting examples in $A$-site ordered quadruple perovskite family. The former shows a large low-field magnetoresistance response with the absence of $Mn^{3+}$-O-$Mn^{4+}$ double-exchange interactions, since only a single $Mn^{4+}$ valence state occurs at the $B$ site [19]. The later displays heavy Fermion behavior since the $d_{xy}$ orbital of the $Cu^{2+}$ strongly hybridizes with the Ir 5$d$ orbitals via the O-$p$ orbitals, so that the local moments of the Cu are incorporated in the itinerant 5$d$ electrons of the Ir [20]. This coupling leads to the disappearance of long-range magnetic ordering of $Cu^{2+}$ ions and strongly enhanced electronic specific heat. Recently, a few $AA'_3B_2B'_2O_{12}$-type quadruple perovskites with orderly distributed 3$d$/5$d$ transition metals at the $B/B'$ sites have been discovered [21-23]. One finds that the introduction of 5$d$ electrons can significantly change the magnetic and electrical transport properties. On account of the interesting properties of CaCu$_3$Mn$_4$O$_{12}$ and CaCu$_3$Ir$_4$O$_{12}$, in this study, we prepare a 3$d$-Mn and 5$d$-Ir hybridized quadruple perovskite oxide CaCu$_3$Mn$_2$Ir$_2$O$_{12}$ (CCMIO) by using high pressure and high temperature conditions. The crystal structure, charge state, magnetism and electrical transport properties are studied in detail.

**2. Experiment details**

The compound of CaCu$_3$Mn$_2$Ir$_2$O$_{12}$ was synthesized as black polycrystalline powders.

Highly pure (> 99.9%) CaO, CuO, MnO and Ir powders with a 1:3:2:2 mole ratio were used as starting materials, and excess $KClO_4$ was applied as an oxidizing agent. These reagents were thoroughly mixed in an agate mortar within an argon-filled glovebox, and then sealed into a platinum capsule with 4.0 mm in diameter and length. The capsule was treated at 5.5 GPa and 1673-1773 K for 60 min on a cubic anvil-type high-pressure apparatus. The sample was quenched to room temperature (RT) once the heat treatment was finished, and then the pressure was gradually released. The product was washed using deionized water to exclude the residual KCl. In addition, $CaCu_3Ti_4O_{12}$ was prepared as described in ref. [24], and $Cu_2O$ single crystal was purchased from MaTeck-Material-Tenchnologie & Kristalle GmbH. These two oxides were used as X-ray absorption references.

The sample quality and crystal structure were identified by powder X-ray diffraction (XRD) on a Huber diffractometer with Cu $K_{\alpha 1}$ radiation ($\lambda$ = 1.5406 Å). The diffraction angle (2θ) varies from 10 to 100° with 0.005° per step. The selected area electron diffraction (SAED) was performed at RT along [110] zone axis by a Philips-CM200 transmission electron microscope with a field emission gun operated at 200 keV. The structure refinement was performed by the Rietveld method using the GSAS program [25]. Soft X-ray absorption spectroscopy (XAS) measurements of Cu-$L_{2,3}$ and Mn-$L_{2,3}$ edges were carried out at beamline BL11A of the National Synchrotron Radiation Research Center (NSRRC) in Taiwan using the total electron yield mode. The Ir-$L_3$ spectra were measured at beamline BL07A of the NSRRC in transmission geometry. The field dependence of the isothermal magnetization ($M$) and temperature-dependent dc magnetic susceptibility ($\chi_{dc}$) were measured using a commercial superconducting quantum interference device magnetometer (MPMS3, Quantum Design). The temperature dependence of the ac magnetization ($M'$) and electrical transport were measured on a physical property measurement system (PPMS7, Quantum Design).

**3. Results and discussions**

The crystal structure of CCMIO is characterized by powder XRD at room temperature. Figure 1(b) shows the XRD pattern as well as the structure refinement result. The Rietveld analysis shows that the compound possesses an *A*-site ordered perovskite structure with space group *Im*-3. It means that the Ca and Cu are 1:3 ordered at the *A* and *A'* sites, respectively, whereas the Mn and Ir are randomly distributed at the *B* site. To confirm the Mn/Ir disorder, the SAED pattern was measured along with the [110] zone axis. As shown in the inset of figure 1(b), one cannot find any diffraction spots with $h + k + l$ = odd such as the (111) and (311) spots, revealing the absence of Mn/Ir long-range order at the B site, in agreement with the XRD result. The refined structure parameters are listed in Table 1. In comparison, the lattice parameter of CCMIO (7.4106 Å) is located between those of $CaCu_3Ir_4O_{12}$ (7.4738 Å) [20] and $CaCu_3Mn_4O_{12}$ (7.2410 Å) [19], suggesting that the Mn and Ir form a solid solution. According to the refined Cu-O bond length, the bond valence sum (BVS) calculations show the valence state of Cu to be +2.09, indicating the presence of a $Cu^{2+}$ state with square-planar coordination. Because of the random Mn/Ir arrangement, we cannot obtain the valence states for these two transition metals by BVS calculations.

The oxidation states of the Cu, Mn and Ir ions in CCMIO were investigated by X-ray absorption at the Cu-$L_{2,3}$, Mn-$L_{2,3}$ and Ir-$L_3$ edges, respectively. Figure 2(a) shows the Cu-$L_{2,3}$ XAS of CCMIO together with those of Cu$_2$O as a Cu$^+$ ($3d^{10}$) reference [26] and CaCu$_3$Ti$_4$O$_{12}$ as a Cu$^{2+}$ ($3d^9$) reference [27] in a quadruple perovskite structure. Typically, for a $3d^9$ Cu$^{2+}$ charge state, a strong absorption peak appears around 930.0 eV due to the $d^9 \rightarrow \underline{c}d^{10}$ transition [26], where $\underline{c}$ indicates a hole in the Cu $2p$ core. Obviously, the Cu-$L_{2,3}$ spectrum of CCMIO exhibits a very strong single peak at the same energy position as that of CaCu$_3$Ti$_4$O$_{12}$ revealing the formation of a Cu$^{2+}$ valence state in CCMIO with a Cu$^{2+}$O$_4$ square-planar local environment. In comparison, the main peak of Cu$_2$O is shifted to higher energy by about 2.5 eV corresponding to a $d^{10} \rightarrow \underline{c}d^{10}s^1$ transition [26] of the fully occupied $3d^{10}$ configuration in Cu$^+$. This does not occur in CCMIO, thus excluding the Cu$^+$ state. Figure 2(b) shows the Mn-$L_{2,3}$ XAS spectra of CCMIO with those of the reference oxides including the tetravalent La$_2$MnCoO$_6$ (Mn$^{4+}$, from ref. 28) and the trivalent LaMnO$_3$ (Mn$^{3+}$, from ref. 28) with similar MnO$_6$ octahedral coordination. Compared with LaMnO$_3$, the main absorption peaks of CCMIO move towards higher energy by more than 1 eV. They have however the same energy position and similar spectral features as those of the Mn$^{4+}$ reference La$_2$MnCoO$_6$. This finding indicates the presence of a Mn$^{4+}$ ($3d^3$) state in CCMIO. The difference at the shoulder of the Mn $L_3$ peak might be due to the slightly different charge transfer and ligand field effects between La$_2$MnCoO$_6$ and CCMIO. The observed Mn$^{4+}$ state in CCMIO is surprising considering the Mn$^{3+}$ state in the Sr$_3$Ru$_{2-x}$Mn$_x$O$_7$ [29].

The Ir-$L_3$ XAS spectra of CCMIO, La$_2$CoIrO$_6$ as an Ir$^{4+}$ reference (from ref. 30) and $Sr_2FeIrO_6$ as an $Ir^{5+}$ reference (from ref. 31) are shown in figure 2(c). The Ir-$L_3$ of CCMIO is shifted by more than 1 eV to lower energy with respect to that of $Ir^{5+}$ reference $Sr_2FeIrO_6$, but locates at the same energy as that of the $Ir^{4+}$ reference $La_2CoIrO_6$ indicating the formation of $Ir^{4+}$ valence state. Therefore, the XAS measurements confirm the charge combination to be $CaCu^{2+}_3Mn^{4+}_2Ir^{4+}_2O_{12}$. Usually, for two cations to form an ordered distribution in a perovskite structure, the charge difference between them is not less than two [32]. The identical charge states between Mn and Ir therefore explains the B-site disorder in CCMIO. In particular, for most quadruple perovskite oxides with hybridized 3d-5d elements at the B site, the 3d-5d charge discrepancy is larger than two and then they form an ordered distribution, as exampled by $CaCu_3Fe_2Re_2O_{12}$, $CaCu_3Fe_2Os_2O_{12}$, and $NaCu_3Fe_2Os_2O_{12}$ [21, 22, 33]. The current CCMIO therefore provides an exception where the 3d and 5d elements have exactly the same charge state.

The magnetism of CCMIO was studied by both dc and ac magnetizations. Figure 3(a) shows the temperature dependence of zero-field-cooling (ZFC) and field-cooling (FC) dc magnetic susceptibility curves measured at different fields. The ZFC and FC curves separate from each other at a critical temperature $T_{SG} \approx 49$ K at 0.01 T. Moreover, the $T_{SG}$ systematically shifts towards lower temperatures with increasing magnetic field, accompanied by a gradual reduction in the intensity of $\chi_{dc}$ and a broadening of the cusp around $T_{SG}$. These features are indicative of a spin glass transition. Figure 3(b) shows the field-dependent magnetization curves measured at different temperatures. At higher temperatures like 200 K, the linear magnetization behavior is observed, as expected for the paramagnetism. At 100 K, the magnetization slightly deviates from the linear dependence on field. It may suggest the formation of some short-range FM interactions. At 2 K, one can find remarkable magnetic hysteresis with a ~2 T coercive field, but the magnetization is still not saturated with fields up to 7 T. These observations are in good agreement with the occurrence of spin glass. Above 240 K, the susceptibility data can be fitted using the modified Curie-Weiss (CW) law $\chi(T) = \chi_0 + C/(T-\theta_w)$, where $\chi_0$ is the temperature independent term containing the core diamagnetism and Van Vleck paramagnetism. The inset of figure 3(a) shows the fitting results, which give a Curie constant $C = 4.75$ emu·K·mol$^{-1}$·Oe$^{-1}$, a Weiss temperature $\theta_w = 58.5$ K, and a $\chi_0 = 3.5\times10^{-3}$ emu·mol$^{-1}$·Oe$^{-1}$. According to the Curie constant, the effective magnetic moment is calculated to be $\mu_{eff} = (8C)^{1/2} = 6.17$ $\mu_B$/f.u. If one only considers the spin-only contribution of Cu$^{2+}$ and Mn$^{4+}$ ions, the effective moment in theory is 6.24 $\mu_B$/f.u., which is very close to the fitting one. Therefore, compared to the 3$d$ Cu$^{2+}$ and Mn$^{4+}$, the 5d Ir$^{4+}$ may play a minor role on the magnetism of CCMIO. Actually, there is no long-range magnetic order occurring in the isostructural compound CaCu$_3$Ir$_4$O$_{12}$ [20].

To further clarify the spin glass behavior of CCMIO, ac magnetization was measured at different frequencies by spanning several orders in magnitude. As shown in figure 4(a), the ac magnetization shows a cusp at a spin freezing temperature $T_f$. Moreover, with increasing frequency, the cusp shifts toward higher temperatures with a reduction in magnitude. These features provide another convincing evidence for the occurrence of spin glass. We can characterize the shift of $T_f$ with frequency using a physical parameter $\kappa = \Delta T_f/[T_f\Delta\ln(\omega)]$, where the spin freezing temperature $T_f$ is determined from the maximum of $M'$, $\omega$ is the ac frequency, and $\Delta T_f$ and $\Delta\ln(\omega)$ represent the differences for different frequencies. The $\kappa$ is calculated to be about $0.8(1)\times10^{-2}$ for CCMIO. The value of $\kappa$ can offer a criterion for distinguishing a spin glass from a super-paramagnet since the later usually has a larger value of $\kappa$. For example, $\kappa$ is 0.28 for the typical super-paramagnet compound a-(Ho$_2$O$_3$) (B$_2$O$_3$) [34] and 0.1 for La$_{0.67}$Sr$_{0.33}$MnO$_3$ nanoparticles [35]. The smaller κ value of CCMIO indicates that the compound should be a spin glass system rather than a super-paramagnet, as reported in other spin glasses like the magnetic B$_{12}$ cluster compound HoB$_{22}$C$_2$N [36]. In addition, the frequency dependence of $T_f$ can be described by a conventional dynamical slowing down model for a three dimension spin glass system with the function $\tau/\tau_0 = [(T_f - T_{SG})/T_{SG}]^{-zv}$ [37]. Here, $\tau_f = 1/\omega$ corresponds to the maximum relation time of the system at $T_f$, $\tau_0$ is the intrinsic relaxation time, and $zv$ is the dynamic exponent. Figure 3(b) shows the plot of $\tau_f$ vs $T_f$ and the fitting result, yielding $zv = 7.67$ and $\tau_0 = 8.22 \times 10^{-13}$ s. The fitted value of $\tau_0$ is in accordance with the typical values of $10^{-11}$-$10^{-13}$ s obtained for many canonical spin glasses such as CuMn [34] and Eu$_{0.5}$Sr$_{1.5}$MnO$_4$ [38].

The spin glass behavior of CCMIO can be attributed to the competing FM and AFM interactions. According to the Goodenough-Kanamori rules [39, 40], for a Mn$^{4+}$ ion with 3$d^3$ configuration in a perovskite octahedral crystal field, the Mn$^{4+}$-O-Mn$^{4+}$ AFM superexchange interaction is expected to occur as observed in CaMnO$_3$ [41]. On the other hand, in *A*-site ordered perovskites, the presence of a net ferromagnetic component is possible due to the Cu$^{2+}$(↑)Mn$^{4+}$(↓) ferrimagnetic coupling, as reported in CaCu$_3$Mn$_4$O$_{12}$ [19]. In addition, for the 5$d$ Ir$^{4+}$, the strong spin-orbital coupling splits the $t_{2g}$ level into a fully occupied $j_{eff} = 3/2$ band and a singly occupied $j_{eff} = 1/2$ state, resulting in a total $J_{eff} = 1/2$ pseudospin state. An AFM interaction is often shown in $J_{eff} = 1/2$ systems, such as Sr$_2$IrO$_4$ and La$_2$CoIrO$_6$ [42, 43]. As for the spin interaction between the *B*-site Mn$^{4+}$ with half-filled $t_{2g}$ orbitals and Ir$^{4+}$ with half-filled $j_{eff} = 1/2$ band is concerned, it most probably leads to Mn$^{4+}$(↑)Ir$^{4+}$(↓) ferrimagnetic correlation as discussed in another 3$d$/5$d$ hybridized system Pb$_2$FeOsO$_6$ [44]. Therefore, a series of complex and competing FM and AFM interactions are responsible for the spin glass behavior of CCMIO.

The temperature dependent resistivity and magnetoresistance effects are measured to characterize the electrical transport properties of CCMIO. As shown in Figure 5(a), the resistivity measured at 300 K is only 15.6 mΩ·cm, which is slightly larger than that observed in $CaCu_3Ir_4O_{12}$ with metallic conductivity arising from the itinerant electronic behavior of $Ir^{4+}$ ions. In CCMIO, although the resistivity lightly increases with decreasing temperature, taking into account the polycrystalline nature where the extrinsic grain boundaries play a role on electrical transport, the weak temperature dependence of resistivity (especially at higher temperatures such as above 100 K) as well as the disordered perovskite structure with itinerant $Ir^{4+}$ state may imply a (bad) metallic electrical behavior. As presented in Figure 6(b), CCMIO shows negative MR effects at different temperatures. Lowering temperature can enhance the MR effects. At 10 K and 7 T, the absolute value of MR is about 10.3%, which is considerably less than that observed in the isostructural compound $CaCu_3Mn_4O_{12}$ due to the decrease of $Mn^{4+}$ component.

**4. Conclusions**

In summary, an Ir-containing quadruple perovskite oxide $CaCu_3Mn_2Ir_2O_{12}$ was prepared at 5.5 GPa and 1673-1773 K. The Rietveld analysis based on the powder XRD data reveals an *A*-site ordered perovskite structure with space group *Im*-3, in which the Mn and Ir ions are disorderly distributed at the *B* site. The valence states and local environment of Cu, Mn and Ir were investigated by XAS techniques, giving $Cu^{2+}O_4$ square-planar coordination at the *A'* site and $Mn^{4+}/Ir^{4+}O_6$ octahedral coordination at the *B* site, resulting in the charge combination of $CaCu^{2+}_3Mn^{4+}_2Ir^{4+}_2O_{12}$. In magnetism, both dc and ac magnetization measurements indicate a spin glass transition around a spin freezing temperature of 49 K. Moreover, the spin glass behavior can be well described by a conventional dynamic slowing down model. In comparison with the isostructural compound $CaCu_3Mn_4O_{12}$, the MR effect of the current CCMIO is reduced considerably owing to the introduction of Ir.

**Acknowledgments**

This work was supported by the National Key R&D Program of China (Grant No. 2018YFA0305700, 2018YFE0103200), the National Natural Science Foundation of China (Grant No. 51772324, 11574378), and the Chinese Academy of Sciences (Grant No. QYZDB–SSW–SLH013, GJHZ1773). The research in Dresden was partially supported by the DFG through SFB 1143. We acknowledge the support from the Max Planck-POSTECH-Hsinchu Center for Complex Phase Materials.

Table 1. Structure parameters and BVS results of CCMIO refined from XRD pattern at room temperature.

| parameter | CCMIO |
|---|---|
| a (Å) | 7.41060(1) |
| $O_y$ | 0.1732(4) |
| $O_z$ | 0.3034(4) |
| $U_{iso}$ for Ca (100×Å$^2$) | 1.19(9) |
| $U_{iso}$ for Cu (100×Å$^2$) | 0.47(2) |
| $U_{iso}$ for Mn/Ir (100×Å$^2$) | 2.45(1) |
| $U_{iso}$ for O (100×Å$^2$) | 0.72(8) |
| $d_{Cu-O}$ (×4) | 1.94203 |
| $d_{B-O}$ (×6) | 1.97846 |
| ∠B-O-B | 139.0(1) |
| ∠Cu-O-B | 109.9(1) |
| BVS (Cu) | 2.09 |
| $R_{wp}$ (%) | 4.95 |
| $R_p$ (%) | 3.25 |

[a]Crystal data: space group *Im*-3 (No. 204); atomic sites: Ca 2a (0, 0, 0); Cu 6b (0, 0.25, 0.25); B (Mn/Ir) 8c (0.25, 0.25, 0.25); O 24h (0, *y*, *z*). The BVS values ($V_i$) were calculated using the formula $V_i = \sum_j S_{ij}$, and $S_{ij} = \exp[(r_0 - r_{ij})/0.37]$. The value of $r_0$ = 1.679 for Cu and 4 coordinated oxygen atoms were used.

**Figures and figure captions**

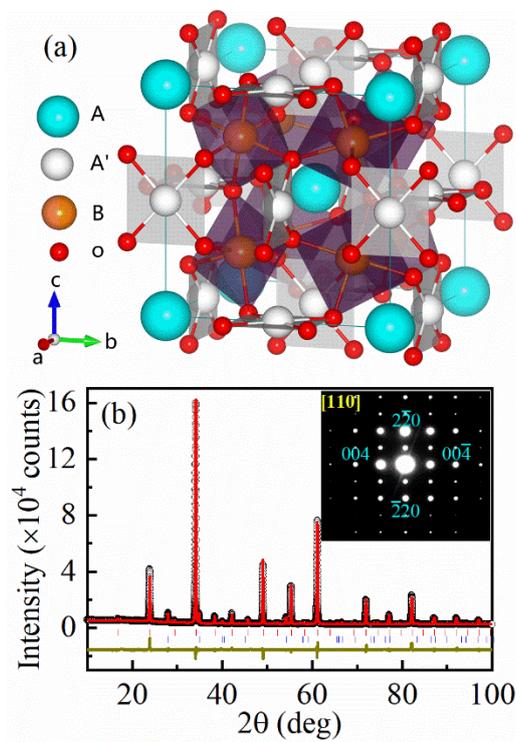

Figure 1. (a) Crystal structure of *A*-site ordered quadruple perovskite $AA'_3B_4O_{12}$ with space group *Im*-3. (b) XRD pattern and Rietveld refinement result for $CaCu_3Mn_2Ir_2O_{12}$. Observed (black circle), calculated (red line) and difference (dark yellow line) values are shown together with the allowed Bragg reflection (above ticks). The lower ticks present a small amount of impurity phase $IrO_2$ (~4.9 wt%). The inset in panel b shows a SAED pattern along the [110] zone axis.

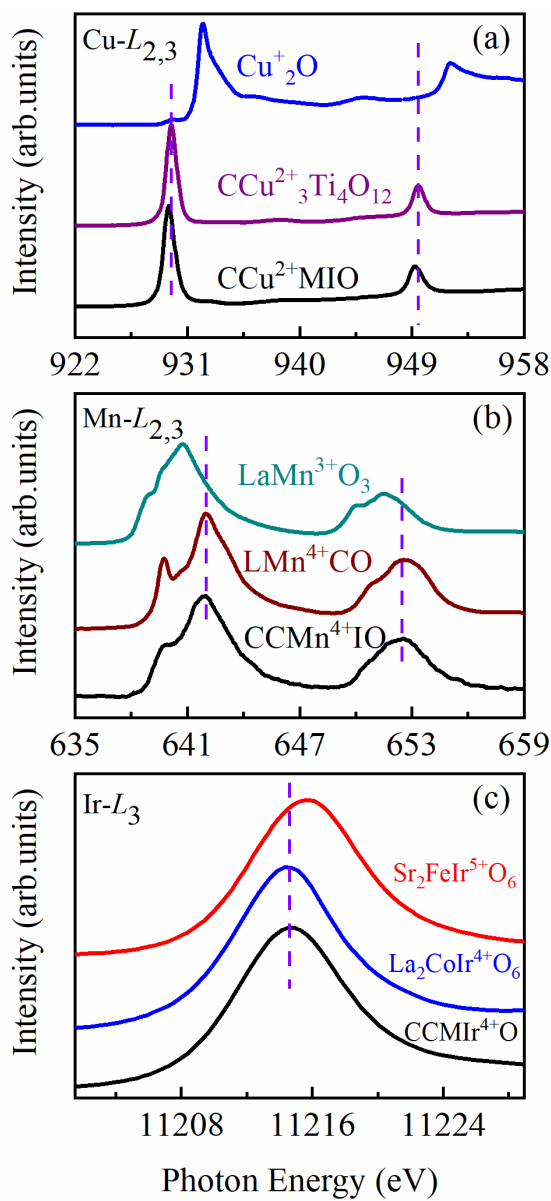

Figure 2. XAS of (a) Cu-$L_{2,3}$ edges of CCMIO and the references CaCu$_3$Ti$_4$O$_{12}$ and Cu$_2$O, and (b) Mn-$L_{2,3}$ edges of CCMIO and the references La$_2$MnCoO$_6$ (LMCO, from ref. 28) and LaMnO$_3$ (from ref. 28), and (c) Ir-$L_3$ edges of CCMIO and the references La$_2$CoIr$^{4+}$O$_6$ (from ref. 30) and Sr$_2$FeIr$^{5+}$O$_6$ (from ref. 31).

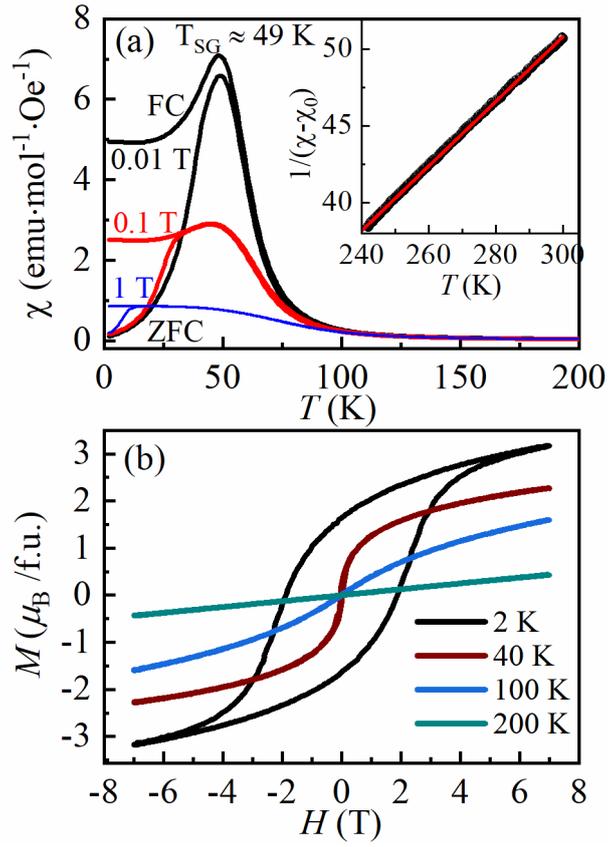

Figure 3. (a) dc magnetic susceptibility of CCMIO measured at different magnetic fields. The inset shows the Curie-Weiss fitting in 240-300 K. (b) Isothermal magnetization measured at selected temperatures.

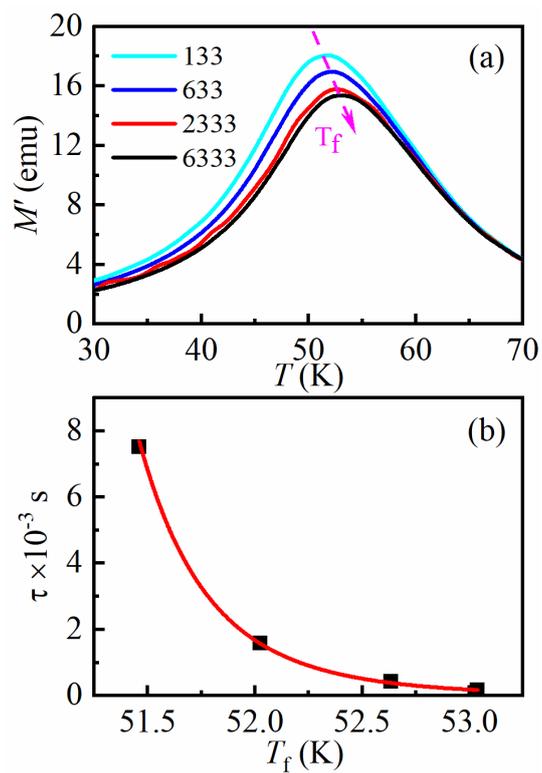

Figure 4. (a) Temperature dependence of ac magnetic susceptibility measured at different frequencies. (b) The fitted results based on the dynamic slowing down model as described in text.

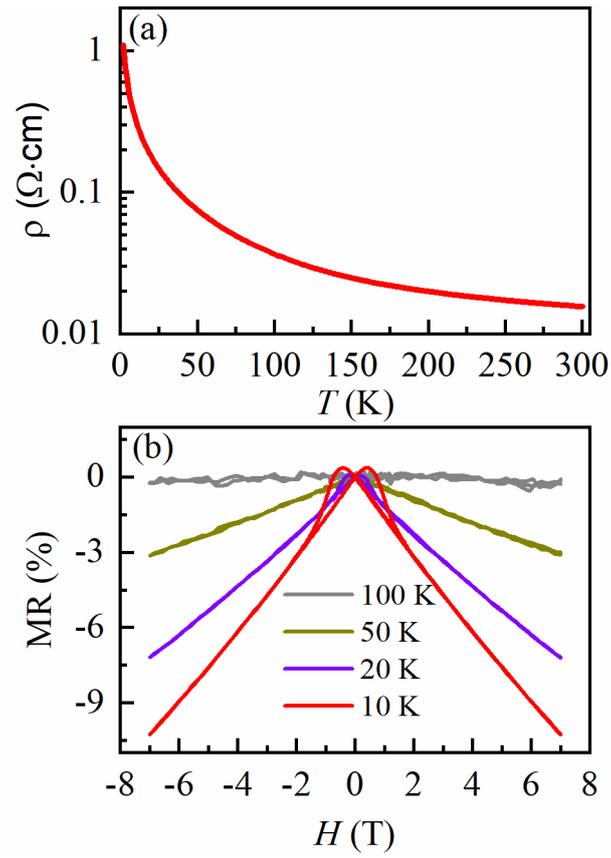

Figure 5. (a) Temperature dependence of resistivity measured on heating at zero field for CCMIO. (b) Magnetoresistance as a function of magnetic field measured at different temperatures.